\documentclass[onecolumn,aps,showpacs,nofootinbib,superscriptaddress]{revtex4}
\usepackage{graphicx}
\usepackage{amsmath}
\usepackage{amssymb}
\usepackage{lscape}
\usepackage{color}
\usepackage{bm}
\usepackage{dsfont}
\usepackage{afterpage}
\usepackage{float}
\usepackage[colorlinks=true, linkcolor=red, citecolor=blue]{hyperref}
\def\be{\begin{equation}}
\def\ee{\end{equation}}
\def\bea{\begin{eqnarray}}
\def\eea{\end{eqnarray}}

\renewcommand{\a}{{A}}
\newcommand{\h}{\mathcal{H}}

\begin{document}
\title{Large-scale stable interacting dark energy model: Cosmological perturbations and observational constraints}

\author{Yun-He Li}
\affiliation{Department of Physics, College of Sciences,
Northeastern University, Shenyang 110819, China}
\author{Xin Zhang\footnote{Corresponding author.}}
\email{zhangxin@mail.neu.edu.cn} \affiliation{Department of Physics,
College of Sciences, Northeastern University, Shenyang 110819,
China} \affiliation{Center for High Energy Physics, Peking
University, Beijing 100080, China}

\begin{abstract}
Dark energy might interact with cold dark matter in a direct, nongravitational way. However, the usual interacting dark energy models (with constant $w$) suffer from some catastrophic difficulties. For example, the $Q\propto\rho_{\rm c}$ model leads to an early-time large-scale instability, and the $Q\propto\rho_{\rm de}$ model gives rise to the future unphysical result for cold dark matter density (in the case of a positive coupling). In order to overcome these fatal flaws, we propose in this paper an interacting dark energy model (with constant $w$) in which the interaction term is carefully designed to realize that $Q\propto\rho_{\rm de}$ at the early times and $Q\propto\rho_{\rm c}$ in the future, simultaneously solving the early-time superhorizon instability and future unphysical $\rho_{\rm c}$ problems. The concrete form of the interaction term in this model is $Q=3\beta H \frac{\rho_{\rm{de}}\rho_{\rm{c}}}{\rho_{\rm{de}}+\rho_{\rm{c}}}$, where $\beta$ is the dimensionless coupling constant. We show that this model is actually equivalent to the decomposed new generalized Chaplygin gas (NGCG) model, with the relation $\beta=-\alpha w$. We calculate the cosmological perturbations in this model in a gauge-invariant way and show that the cosmological perturbations are stable during the whole expansion history provided that $\beta>0$. Furthermore, we use the Planck data in conjunction with other astrophysical data to place stringent constraints on this model (with eight parameters), and we find that indeed $\beta>0$ is supported by the joint constraint at more than 1$\sigma$ level. The excellent theoretical features and the support from observations all indicate that the decomposed NGCG model deserves more attention and further investigation.
\end{abstract}

 \maketitle

\section{Introduction}\label{sec:intro}

The current Universe is dominated by two dark sectors, namely, dark energy (DE) and dark matter (DM), which is supported by recent astronomical
observations \cite{Riess98,Tegmark04,Spergel03}. However, we still know little about their nature and can only indirectly detect them via their gravitational effects. This provides us more room to study the possibility that there exists some direct, nongravitational interaction between DE and DM. Such a possible interaction can help solve or alleviate several theoretical problems of DE, such as the cosmic coincidence problem \cite{intde1}, the cosmic doomsday problem led by phantom \cite{intde2}, and the cosmic age problem caused by old quasars \cite{intde3}. Besides, DE can also exert a nongravitational influence on DM by dark sector interaction, inducing new features to structure formation, such as new large scale bias \cite{Amendola:2001rc} and violation of the weak equivalence principle for DM \cite{Bertolami:2007zm,Koyama:2009gd}. Thus, it is very meaningful to study such an interaction between DE and DM.

The dark sector interaction in the background evolution can be characterized by adding an interaction term $Q$ to the energy balance equations of DE and DM, i.e.,
\begin{eqnarray}
 \label{rhodedot} \dot{\rho}_{\rm{de}} &=& -3H(1+w)\rho_{\rm{de}}+ Q_{\rm{de}}, \\
 \label{rhocdot} \dot{\rho}_{\rm{c}} &=& -3H\rho_{\rm{c}}+ Q_{\rm{c}},~~~~~~Q_{\rm{de}}=-Q_{\rm{c}}=Q,
\end{eqnarray}
where $\rho_{\rm{de}}$ and $\rho_{\rm{c}}$ are the energy densities of DE and DM (here, specifically, cold dark matter), respectively,
$H=\dot{a}/a$ is the Hubble expansion rate, and a dot denotes the derivative with respect to the cosmic time $t$, $a$ is the scale factor of
the Friedmann-Robertson-Walker (FRW) universe, and
$w=p_{\rm{de}}/\rho_{\rm{de}}$ is the equation of state (EOS) parameter of DE. Due to the fact that the knowledge about the micro-origin of the dark sector interaction is absent, one has to propose the interacting DE models by writing down the possible forms of $Q$ by hand. So far, lots of phenomenological forms for $Q$ have been put forward \cite{Amendola:1999qq,Billyard:2000bh,Zimdahl:2001ar,Farrar:2003uw,Chimento:2003iea,Olivares:2005tb,Koivisto:2005nr,Sadjadi:2006qp,Guo:2007zk,Zhang:2007uh,Boehmer:2008av,intde5,Quartin:2008px,
Pereira:2008at,Quercellini:2008vh,Valiviita:2008iv,Bean:2008ac,Chongchitnan:2008ry,Corasaniti:2008kx,CalderaCabral:2008bx,Jackson:2009mz,He:2009pd,He:2010ta,Cai:2009ht,Li:2011ga,Clemson:2011an,wangbin12,wangbin13a,wangbin13b}. Among them, the models with $Q\propto H\rho$ and $Q\propto\rho$ (with $\rho$ either the energy density of DE/DM or the sum of the two) are widely studied. For the dynamical analyses of the systems in such interacting DE models, including the situations of the fixed points and their stability, see, e.g., Refs.~\cite{dyn1,dyn2,dyn3,dyn4}.

In recent years, it has been found that the interacting DE models may suffer from a large-scale instability at the early times if the EOS of DE is taken to be a constant. In Ref.~\cite{Valiviita:2008iv}, the authors gave a detailed investigation on the perturbation evolutions for the three interacting DE models, $Q=\Gamma\rho_{\rm{c}}$, $Q=\gamma H\rho_{\rm{c}}$, and $Q=\gamma H(\rho_{\rm{c}}+\rho_{\rm{de}})$, where $\Gamma$ and $\gamma$ are coupling constants. They found that all of them cannot give stable cosmological perturbations at the early times if $w=\rm{const}$ and $w>-1$, while if $w<-1$, the perturbations are stable. However, it is well known that the $w<-1$ case will lead to another instability of our universe in a finite future, and thus usually it is not considered as an acceptable case. Note that throughout the paper we do not consider the case of $w<-1$. After this, the instability in the interacting DE models was reexamined~\cite{He:2008si}. It is found that the instability depends on the type of the interacting DE model: if $Q=3\beta H\rho_{\rm{de}}$, stable cosmological perturbations could be given, provided that $\beta>0$. The same stability condition was also pointed out in Ref.~\cite{Clemson:2011an} for the $Q=\Gamma\rho_{\rm{de}}$ model. Thus, it seems that the case with $Q$ proportional to $\rho_{\rm{de}}$ and with a positive coupling constant provides us with the most acceptable interacting DE model.\footnote{It was shown that a time-dependent $w$ can help solve the early-time instability problem \cite{Majerotto:2009np}. However, a time-dependent $w$ will introduce at least one more free parameter.} Indeed, a positive coupling is favored by observations; see, e.g., Refs.~\cite{Clemson:2011an,Gavela:2009cy,Gavela:2010tm,Salvatelli:2013wra}. However, this does not mean that there is no problem in this interacting DE model. Actually, a positive coupling in the model with $Q$ proportional to $\rho_{\rm{de}}$ will lead to a negative value of $\rho_{\rm{c}}$ in the future. For example, for the $Q=3\beta H\rho_{\rm{de}}$ case, $\rho_{\rm{c}}=\rho_{\rm{c0}}a^{-3}(1+r-ra^{3\beta-3w})$ for a constant $w$, where $r\equiv\beta\rho_{\rm{de0}}/[\rho_{\rm{c0}}(\beta-w)]$ and the subscript ``0'' denotes the present value of the corresponding quantity, and one can check that $\rho_{\rm{c}}<0$ after $a\simeq1.35$ if choosing $\beta=0.1$, $w=-0.98$, and $\rho_{\rm{c0}}/\rho_{\rm{de0}}=0.36$. This nonphysical result arises from the fact that a positive coupling results in energy transfer from DM to DE, and the interaction term $Q\propto H\rho_{\rm{de}}$ (or $Q\propto \rho_{\rm{de}}$) exacerbates this energy transfer in the DE dominated future. Note that the models with $Q$ proportional to $\rho_{\rm{c}}$ do not have this problem.

In short, for the interacting DE models with constant $w$, the knowledge acquired from the above discussions can be briefly summarized as: $Q\propto\rho_{\rm{c}}$ leads to a large-scale instability at the early times, and $Q\propto\rho_{\rm{de}}$ gives rise to a negative $\rho_{\rm{c}}$ in the future. Therefore, it is fairly natural to design an interacting DE model (with constant $w$) in which the interaction term $Q$ is proportional to $\rho_{\rm{de}}$ at the early times and proportional to $\rho_{\rm{c}}$ in the future. We expect that in this model the cosmological perturbations will always be stable during the whole expansion history of the universe and the negative value of $\rho_{\rm{c}}$ will not occur. We shall show that such a reasonable interacting DE model can emerge from an existing unified dark fluid scenario, namely, the new generalized Chaplygin gas (NGCG) scenario~\cite{Zhang:2004gc}.

Let us consider the interaction form
\begin{equation}\label{interaction}
Q=3\beta H \rho_{\rm{de}}R_{\rm{c}},
\end{equation}
with $R_{\rm{c}}\equiv\frac{\rho_{\rm{c}}}{\rho_{\rm{de}}+\rho_{\rm{c}}}$.
It is clear to see that this form of $Q$ satisfies the above conditions.
The interacting DE model with this form of $Q$ and a constant $w$ can be obtained from the NGCG scenario~\cite{Zhang:2004gc} by setting $\beta=-\alpha w$ with $\alpha$ the NGCG model parameter (for the detailed derivation, see Appendix~\ref{appendixNGCG}). So, this interacting DE model is equivalent to a decomposed NGCG model. Actually, a decomposed generalized Chaplygin gas (GCG) model has been discussed~\cite{Wang:2013qy}.
In Ref.~\cite{Wang:2013qy}, two covariant interaction models are investigated---(1) the so called ``barotropic model'' in which the covariant interaction form is proportional to the gradient of the local DM density and (2) the ``geodesic model'' where the energy-momentum transfer is parallel to the DM four-velocity.
The ``geodesic model'' is widely studied in the literature; in this model there is no momentum transfer in the rest frame of DM.
When the cosmological perturbations are absent, the ``geodesic model'' naturally goes back to the background interaction form, while for the  ``barotropic model'' how
to return back is obscure. Therefore, in this work, we only consider the  ``geodesic model''.
Note also that the decomposed GCG model~\cite{Wang:2013qy} describes DM interacting with the vacuum energy ($w=-1$), and thus the perturbation of DE is always zero in the DM-comoving frame within the ``geodesic model''. In our work, we focus on the $w={\rm const}$ case, for which one must seriously treat the DE perturbation, as there may exist the large-scale instability mentioned above. We shall show that the model with $Q$ given by Eq.~(\ref{interaction}) and a constant $w$ (or, the decomposed NGCG model) is a reasonable, large-scale stable interacting DE model.

Using Eqs.~(\ref{rhodedot})--(\ref{interaction}), we can obtain the background energy densities of DE and DM,
\begin{eqnarray}
  \label{eq:ngcgde}\rho_{\rm{de}} & = & \rho_{\rm{de0}} a^{-3(1+w-\beta)}\left[R_{\rm{c0}}+(1-R_{\rm{c0}})a^{-3(w-\beta)}\right]^{\frac{\beta}{w-\beta}},  \\
  \label{eq:ngcgc}\rho_{\rm{c}} & = &\rho_{\rm{c0}} a^{-3}\left[R_{\rm{c0}}+(1-R_{\rm{c0}})a^{-3(w-\beta)}\right]^{\frac{\beta}{w-\beta}}.
\end{eqnarray}
From the above equations, one can clearly see that both the energy densities of DE and DM are always positive from the past to the future no matter what sign of $\beta$ takes, since $0<R_{\rm{c0}}<1$. Thus, this interacting DE model overcomes the flaw that a positive coupling leads to the future nonphysical evolution of $\rho_{\rm{c}}$ in the $Q=\Gamma\rho_{\rm{de}}$ or $Q=\gamma H\rho_{\rm{de}}$ model. Furthermore, we will show that this model can also give stable cosmological perturbations at the early times.

Our paper is organized as follows. In Sec.~\ref{sec:pertur}, we give the general gauge-dependent perturbation equations for the present interacting DE model. Following Ref.~\cite{Gavela:2010tm}, we will consider the perturbation of $H$ in Eq.~(\ref{interaction}) in order to derive the gauge invariant evolution equations. In Sec.~\ref{sec:perturstable}, we discuss the stability of cosmological perturbations using a gauge invariant way. In Sec.~\ref{sec:constraint}, we use the Planck data and other observations to constrain the model. We will show that a positive coupling constant $\beta$ required by the stable perturbations is also favored by the current observations. We will give conclusions in the final section. In our analysis, we only care about the $w\geq-1$ case to avoid future instability of our Universe.

\section{Perturbation equations}\label{sec:pertur}

In this section, we give the general gauge-dependent perturbation equations for the considered interacting DE model. For simplicity, we only consider a flat universe. Extending the result to a nonflat universe is straightforward. We follow the notation of Ref.~\cite{Valiviita:2008iv} and from here on we use the conformal time $\eta$ (defined as $\rm{d}\eta=\rm{d}\it{t}/\it{a}$) as the independent variable instead of the cosmic time $t$. So, the conformal Hubble expansion rate is $\mathcal{H}=Ha$.
The flat FRW metric with scalar perturbations can be written in general as
\begin{equation}
\label{eq:gen gauge} {\rm d} s^2=a^2\Big\{ -(1+2\phi){\rm d}\eta^2+2\partial_iB\,{\rm d}\eta {\rm d}x^i +\Big[(1-2\psi)\delta_{ij}+2\partial_i\partial_j E\Big]{\rm d} x^i{\rm d} x^j\Big\},
\end{equation}
where $\phi$, $B$, $\psi$ and $E$ are the gauge-dependent scalar metric perturbation quantities.
For the given metric~(\ref{eq:gen gauge}), one does not need to modify the linear Einstein equations for the dark sector interaction case, but needs to modify the conservation equations for $A$ fluid,
\begin{equation}
\label{eqn:energyexchange} \nabla_\nu T^{\mu\nu}_{\a } = Q^\mu_{\a}\,, \quad\quad
 \sum_A Q^\mu_{\a } = 0,
\end{equation}
where $Q^\mu_{\a}$ denotes the energy-momentum transfer for $A$ fluid, and $T^{\mu\nu}_{\a }$ is the $A$-fluid energy-momentum tensor,
\begin{equation}
T_{\a\,\nu}^\mu =(\rho_\a+ p_\a)u_\a^\mu u_\nu^\a+p_\a\delta^\mu{}_\nu +\pi^\mu_{\a\,\nu},
\end{equation}
where $\pi^\mu_{\a\,\nu}$ is the $A$-fluid anisotropic stress, and we note that $\rho_\a$ and $p_\a$ contain the contributions of corresponding perturbations $\delta\rho_\a$ and $\delta p_\a$, respectively. The $A$-fluid four-velocity is given by
 \begin{equation}\label{ua}
u^\mu_\a = a^{-1}\Big(1-\phi, ~\partial^i v_\a \Big)\,, \quad\quad
u_\mu^\a = a\Big(-1-\phi, ~\partial_i[ v_\a +B] \Big),
 \end{equation}
with $v_\a$ the $A$-fluid peculiar velocity potential. In our work, we use the $A$-fluid volume
expansion rate $\theta_\a$~\cite{Ma:1995ey},
 \begin{equation}
\theta_\a=-k^2 (v_\a+B),
 \end{equation}
where $k$ is the comoving wave number in the Fourier space.

To complete Eq.~(\ref{eqn:energyexchange}), one needs a covariant energy-momentum transfer form. However, we cannot obtain it from the first principle. In our work, we construct it using the background interaction term~(\ref{interaction}). First, we follow Refs.~\cite{Kodama:1985bj,Malik:2002jb} and split $Q^\a_\mu$ relative to the total four-velocity,
\begin{eqnarray}
Q^{\a }_0 & = & -a\Big[ Q_\a(1+\phi) + \delta Q_\a
\Big],\label{eqn:Qenergy}
\\
Q^{\a }_i & = & a\partial_i\Big( f_\a- Q_\a \frac{\theta}{k^2}\Big),
\label{eqn:Qmomentum}
\end{eqnarray}
where $f_A$ represents the momentum transfer potential and $\theta$ is the total velocity perturbation.
Then, the energy and momentum balance equations for $A$ fluid from Eq.~(\ref{eqn:energyexchange}) are given by \cite{Valiviita:2008iv}
\begin{eqnarray}
&&\delta_\a'+3{\cal H}(c_{{\rm s}\a}^2-w_\a)\delta_\a+(1+w_\a)\theta_\a+ 3{\cal H}\big[3{\cal
H}(1+w_\a)(c_{{\rm s}\a}^2-w_\a)+w_\a' \big] {\theta_\a \over k^2}
\nonumber\\
&&~~~-3(1+w_\a)\psi'+ (1+w_\a)k^2\big(B-E'\big) ={aQ_\a \over
\rho_\a}\left[ \phi-\delta_\a+3{\cal H}(c_{{\rm s}\a}^2-w_\a) {\theta_\a
\over k^2} \right]
+{a\over \rho_\a}\, \delta Q_\a\,,\label{dpa}\\
&& \theta_\a'+{\cal H}\big(1-3c_{{\rm s}\a}^2\big)\theta_\a- {c_{{\rm s}\a}^2
\over (1+w_\a)}\,k^2\delta_\a +{2\over
3a^2(1+w_\a)\rho_\a}\,k^4\pi_\a -k^2\phi
\nonumber \\
&&~~~= {aQ_\a \over (1+w_\a)\rho_\a}\big[ \theta-
(1+c_{{\rm s}\a}^2)\theta_\a \big] -{a\over (1+w_\a)\rho_\a} \,
k^2f_\a\,, \label{vpa}
 \end{eqnarray}
where $\delta_\a=\frac{\delta\rho_A}{\rho_A}$, the prime denotes the derivative with respect to the conformal time $\eta$, and $c_{{\rm s}\a}^2$ is the sound speed of $A$ fluid. For a barotropic fluid, $c_{{\rm s}\a}^2=c_{{\rm a}\a}^2$ with $c_{{\rm a}\a}^2$ the adiabatic sound speed of $A$ fluid defined by $c_{{\rm a}\a}^2\equiv p'_A/\rho'_A$. However, for the DE perturbation, we cannot take $c_{\rm s,de}^2=c_{\rm a,de}^2$, since $c_{\rm a,de}^2=w<0$ leads to instability in the dark energy~\cite{Gordon:2004ez}. So, it is necessary to assume that DE is a nonadiabatic fluid and impose $c_{\rm s,de}^2>0$ by hand. In our work, as usual, we take $c_{{\rm s},{\rm{de}}}^2=1$; this is what is done in the {\tt CAMB} code \cite{camb}.

Next, we calculate $\delta Q$ for our interacting DE model. From Eq.~(\ref{interaction}), we have
\begin{equation}\label{eq:deltaq}
\delta Q=Q\left[\frac{\delta H}{H}+R_{\rm{c}}\delta_{\rm{de}}+(1-R_{\rm{c}})\delta_{\rm{c}}\right].
\end{equation}
Note that here we consider the perturbation of the Hubble parameter, which is indispensable for the gauge invariant equations (\ref{eq.Dc_newnew}) and (\ref{eq.Dx_newnew}) in the next section. That is to say, without the help of the term $\frac{\delta H}{H}$, one cannot get the gauge invariant equations for a dark sector coupling case, if the interaction term $Q$ is proportional to $H$. In Ref.~\cite{Gavela:2010tm}, the authors pointed out this problem and tried to solve it by considering the perturbation of $H$ for the first time. In our work, we follow Ref.~\cite{Gavela:2010tm} and take $\mathcal{K}\equiv\frac{1}{\mathcal{H}}\left[\frac{\theta}{3}-\mathcal{H}\phi-\psi'+\frac{k^2}{3}\left(B-E'\right)\right]$ as the perturbation of $H$. (Note that their notation of the metric perturbations is different from ours; the corresponding relationships are $A=\phi$ and $H_L=-\frac{1}{3}k^2E-\psi$). Substituting Eq.~(\ref{interaction}) into Eq.~(\ref{eq:deltaq}), and taking $\frac{\delta H}{H}=\mathcal{K}$, we have
\begin{equation}\label{eq:deltaqc}
a\delta Q_{\rm{c}}=-a\delta Q_{\rm{de}}=-3\beta\mathcal{H}\rho_{\rm{de}}R_{\rm{c}} \left[\mathcal{K}+R_{\rm{c}}\delta_{\rm{de}}+(1-R_{\rm{c}})\delta_{\rm{c}}\right].
\end{equation}

The momentum transfer potential $f_A$ cannot be derived from the background interaction term~(\ref{interaction}), and one needs to specify it by hand. In the literature, one often chooses it by assuming that the energy-momentum transfer is parallel to the four-velocity of either DM or DE, so that the momentum transfer vanishes either in the DM-rest frame or in the DE-rest frame. In our work, we only focus on the former. We also note that the stability of the cosmological perturbations is independent of the choice of the energy-momentum transfer type; for this point, see Ref.~\cite{Clemson:2011an} and Appendix~\ref{appendixDEV} of this paper. In Appendix~\ref{appendixDEV}, the case with the energy-momentum transfer parallel to the DE four-velocity is also briefly discussed. In this work, as a concrete example, we only analyze in detail the case with the energy-momentum transfer parallel to the DM four-velocity, i.e.,
\begin{equation}
aQ_{\rm{c}}^\mu=-aQ^\mu_{\rm{de}}=-3\beta\mathcal{H} \rho_{\rm{de}}R_{\rm{c}} u_{\rm{c}}^\mu.
\end{equation}
Using Eq.~(\ref{ua}), one can get
\begin{equation}\label{eq:covqc}
aQ^{\rm{c}}_\mu =-aQ^{\rm{de}}_\mu = 3\beta\mathcal{H}a\rho_{\rm{de}}R_{\rm{c}}\Big[ 1+\phi+\mathcal{K}+ R_{\rm{c}}\delta_{\rm{de}}+(1-R_{\rm{c}})\delta_{\rm{c}}\,,~
\partial_i\left(v_{\rm{c}}+B \right) \Big].
\end{equation}
Comparing Eq.~(\ref{eq:covqc}) with Eq.~(\ref{eqn:Qmomentum}),
one can find
\begin{equation}\label{eq:fc}
ak^2f_{\rm{c}}=-ak^2f_{\rm{de}}=3\beta\mathcal{H}\rho_{\rm{de}} R_{\rm{c}}(\theta_{\rm{c}}-\theta).
\end{equation}

Finally, with the help of Eqs.~(\ref{eq:deltaqc}) and (\ref{eq:fc}), $\pi_{\rm{c}}=0=\pi_{\rm{de}}$, and $c^2_{{\rm s}\,\rm{c}}=w_{\rm{c}}=0=w'$. For our interacting DE model, Eqs.~(\ref{dpa}) and
(\ref{vpa}) can be written as
 \bea
\label{eq.delta'de_ourQ} && \delta'_{\rm{de}} + 3\mathcal
H(1-w)\delta_{\rm{de}} + (1+w)\left[\theta_{\rm{de}} +k^2
(B-E')\right] + 9\mathcal H^2(1-w^2)\frac{\theta_{\rm{de}}}{k^2}
-3(1+w)\psi' \nonumber\\
&&~~~~=3\beta\mathcal{H}R_{\rm{c}}\left[\mathcal{K}+(1-R_{\rm{c}})(\delta_{\rm{c}}-\delta_{\rm{de}})
+ 3\mathcal
H (1-w)\frac{\theta_{\rm{de}}}{k^2}+\phi\right], \\
&& \theta'_{\rm{de}} -2 \mathcal H\theta_{\rm{de}}
-\frac{k^2}{(1+w)}\delta_{\rm{de}} - k^2\phi =
\frac{3\beta\mathcal{H}}{1+w}R_{\rm{c}}
\left(\theta_{\rm{c}}-2\theta_{\rm{de}}
\right)\,, \label{eq.theta'de_ourQ}\\
\label{eq.delta'c_ourQ} && \delta'_{\rm{c}} +\theta_{\rm{c}} + k^2 (B-E')
-3\psi' =-3\beta\mathcal{H}(1-R_{\rm{c}})[\mathcal{K}+R_{\rm{c}}(\delta_{\rm{de}}-\delta_{\rm{c}})+\phi]\,, \\
&& \theta'_{\rm{c}} + \mathcal H \theta_{\rm{c}} -k^2\phi = 0\,.
\label{eq.theta'c_ourQ}
 \eea

\section{Large-scale stability and initial conditions for perturbations}\label{sec:perturstable}

As mentioned in Sec.~\ref{sec:intro}, many interacting DE models suffer from the early-time large-scale instabilities. Such instabilities arise from the fact that the nonadiabatic mode soon dominates and leads to rapid growth of curvature perturbation at the early times, even if the adiabatic initial conditions are utilized \cite{Valiviita:2008iv,Majerotto:2009np}. Thus, analyzing such instabilities is closely related to analyzing the initial conditions for cosmological perturbations. In Ref.~\cite{Doran:2003xq}, the authors presented a systematic approach to obtaining the initial conditions of cosmological perturbations in a noninteracting dark sector case using a gauge invariant way. In that approach, the solutions to the perturbation equations of each component are reduced to those of a first order differential matrix equation,
\begin{equation}
\label{eq.dUdlnx_gen} \frac{\mathrm{d} \bm{U}}{\mathrm{d} \ln x} = \mathbf{A}(x) \bm{U}(x),
\end{equation}
where $x=k\eta$, and
\begin{equation}\label{eq.GIvarsVertors}
\bm{U}^T = \left\{ \Delta_{\rm{c}}, \, \tilde{V}_{\rm{c}}, \,
\Delta_\gamma, \, \tilde{V}_\gamma, \, \Delta_{\rm{b}}, \, \Delta_\nu, \,
\tilde{V}_\nu ,\, \tilde{\Pi}_\nu , \, \Delta_{\rm{de}}, \,
\tilde{V}_{\rm{de}} \right\}.
 \end{equation}
Here, the subscripts $\gamma$, $\rm{b}$, and $\nu$ represent photons, baryons, and neutrinos, respectively. $\Delta_A$, $V_A$ and $\Pi_A$ are gauge invariant variables for matters devised by Bardeen \cite{Bardeen}:
\begin{eqnarray}
\label{eq.GI_denandvel} && \Delta_A = \delta_A +
\h^{-1}\frac{\rho_A'}{\rho_A} \psi\,,  \quad \quad \quad V_A =
k^{-1}\theta_A + k(B -E')\,, \quad \quad \quad \Pi_A = \pi_A .
\end{eqnarray}
Note that $\tilde{V}_A$ and $\tilde{\Pi}_A$ in Eq.~(\ref{eq.GIvarsVertors}) are the corresponding rescaled quantities, namely, $\tilde{V}_A=V_A/x$ and $\tilde{\Pi}_A=\Pi_A/x^2$, respectively. Besides the gauge invariant variables of matters, the metric gauge invariant variables $\Phi$ and $\Psi$ are also used, constructed by \cite{Bardeen}
\begin{eqnarray}
\label{eq.PhiandPsi} && \Phi = -\psi + \h(B-E') \,, \quad \quad
\quad
\Psi =  \phi + \h \left( B-E' \right) + \left( B-E'\right)'.
\end{eqnarray}

In Ref.~\cite{Majerotto:2009np}, the authors generalized the analyzing approach in Ref.~\cite{Doran:2003xq} to the dark sector coupling case. In this part, we apply it to our analysis of the perturbation stability and initial conditions for our interacting DE model. First, we rewrite the perturbation equations for each component in terms of the gauge invariant variables. Since we care about the solutions in the early radiation dominated epoch, we can take $\h=\eta^{-1}$. Then, Eqs.~(\ref{eq.delta'de_ourQ})--(\ref{eq.theta'c_ourQ}) become
\bea    \frac{\mathrm{d} \Delta_{\rm{c}}}{\mathrm{d}\ln x}& =& -x^2\tilde{V_{\rm{c}}} -3\beta(1-R_{\rm{c}})\left[R_{\rm{c}}(\Delta_{\rm{de}}-\Delta_{\rm{c}})+\frac{x^2}{3}\tilde{V}\right]\,,
\label{eq.Dc_newnew}\\     \frac{\mathrm{d} \tilde{V}_{\rm{c}}}{\mathrm{d}\ln x} & =& -2\tilde{V_{\rm{c}}} + \Psi\,, \label{eq.Vc_newnew}\\
\frac{\mathrm{d} \Delta_{\rm{de}}}{\mathrm{d}\ln x} &=& 3(w-1)\left\{ \Delta_{\rm{de}} + 3
(1+w)\left( \Psi +\Omega_\nu \tilde{\Pi}_\nu  \right)
+(1+w)\left[ 3 - \frac{x^2}{3(w -1)} \right]
\tilde{V}_{\rm{de}} \right\} \nonumber \\
  && +3\beta R_{\rm{c}} \left[
(1-R_{\rm{c}})(\Delta_{\rm{c}} -\Delta_{\rm{de}}) +3(1-w)\left( \tilde{V}_{\rm{de}}+\Psi +\Omega_\nu \tilde{\Pi}_\nu  \right)
+\frac{x^2}{3}\tilde{V}  \right],
\label{eq.Dx_newnew} \\
\frac{\mathrm{d} \tilde{V}_{\rm{de}}}{\mathrm{d}\ln x}&=& \frac{\Delta_{\rm{de}}}{1+w} +
\tilde{V}_{\rm{de}} +  3 \Omega_\nu \tilde{\Pi}_\nu + 4\Psi  +
3\beta R_{\rm{c}} \frac{\tilde{V}_{\rm{c}} - 2 \tilde{V}_{\rm{de}} - \Omega_\nu \tilde{\Pi}_\nu
- \Psi }{1+w}
    \label{eq.Vx_newnew},
 \eea
with $\Psi$ given by
 \be
\Psi = -\frac{\sum_{A=\rm{c}, \rm{b}, \gamma, \nu, \rm{de}} \Omega_A \left[
\Delta_A + 3 \,(1+w_A) \tilde{V}_A \right]}{\sum_{A=\rm{c}, \rm{b}, \gamma, \nu, \rm{de}} 3 \,(1+w_A) \Omega_A + \frac{2}{3} x^2} -
\Omega_\nu \tilde{\Pi}_\nu \,. \label{eq.Psi}
 \ee
Here, we have used the Einstein equation $\Phi=-\Psi-\Omega_\nu \tilde{\Pi}_\nu$ and defined $\Omega_A\equiv\rho_A/\rho_{\rm{crit}}$ for $A$ fluid with $\rho_{\rm{crit}}$ the critical density of our universe. For other components, they satisfy the same differential equations as those of the uncoupled case, given in Ref.~\cite{Doran:2003xq}.

Next, we give the coefficient matrix $\mathbf{A}(x)$ in Eq.~(\ref{eq.dUdlnx_gen}). At the early times, $x\ll1$, $\mathbf{A}(x)$ can be reduced to a constant matrix $\mathbf{A}_0$, as long as no divergence occurs when $x\rightarrow0$. We can also take $a=\h_0\sqrt{\Omega_{\rm{r0}}}\eta$ and $\Omega_A\simeq\rho_A/\rho_{\rm{r}}$, since the early universe is dominated by radiation; thus we have
 \bea
 && \Omega_{\rm{b}} = \frac{\rho_{\rm{b}}}{\rho_{\rm{r}}} =
\frac{\Omega_{\rm{b0}}}{\Omega_{\rm{r0}}} \, a =
\frac{\Omega_{\rm{b0}}}{\sqrt{\Omega_{\rm{r0}}}}\frac{\h_0}{k} \, x =
\omega_1 \,x \,, \quad \quad \quad \Omega_{\rm{c}} =
\frac{\Omega_{\rm{c0}}R_{\rm{c0}}^{\beta/(w-\beta)}}{\sqrt{\Omega_{\rm{r0}}}}\frac{\h_0}{k} \, x=\omega_2 \, x \,, \nonumber\\
&& \Omega_{\rm{de}} =\frac{\Omega_{\rm{de0}}R_{\rm{c0}}^{\beta/(w-\beta)}}{\Omega_{\rm{r0}}}\left(\frac{\sqrt{\Omega_{\rm{r0}}}\h_0}{k}\right)^{1-3(w-\beta)} \, x^{1-3(w-\beta)}=\omega_3 \, x^{1-3(w-\beta)} ,\nonumber \\
&& \Omega_\nu = \rho_\nu /\rho_{\rm{r}} = R_\nu \,, \quad \quad \quad
\Omega_\gamma = 1- \Omega_{\rm{b}} - \Omega_{\rm{c}} - \Omega_{\rm{de}} -\Omega_\nu
\,. \label{eq.Omegas_expl}
 \eea
Here, for DE and DM, we have used Eqs.~(\ref{eq:ngcgde}) and (\ref{eq:ngcgc}), and neglected the $(1-R_{\rm{c0}})a^{-3(w-\beta)}$ term. Note that there is no divergence term in Eq.~(\ref{eq.Omegas_expl}) when $x\rightarrow0$ under the assumptions $w<-1/3$ and small coupling constant $\beta$ required by observations. Then, the coefficient matrix $\mathbf{A}(x)$ at the zeroth order is given by
\be
\label{eq:A0}
\mathbf{A}_0=
\left(
\begin{array}{cccccccccc}
 0 & 0 & 0 & 0 & 0 & 0 & 0 & 0 & 0 & 0 \\
 0 & -2 & \frac{\mathcal{N}}{4} & \mathcal{N} & 0 &-\frac{R_\nu}{4} & -R_\nu & -R_\nu & 0 & 0 \\
 0 & 0 & 0 & 0 & 0 & 0 & 0 & 0 & 0 & 0 \\
 0 & 0 & \frac{2R_\nu -1}{4} & 2R_\nu -3 & 0 &-\frac{R_\nu}{2} & -2R_\nu & -R_\nu & 0 & 0 \\
 0 & 0 & 0 & 0 & 0 & 0 & 0 & 0 & 0 & 0 \\
 0 & 0 & 0 & 0 & 0 & 0 & 0 & 0 & 0 & 0 \\
 0 & 0 & \frac{\mathcal{N}}{2} & 2\mathcal{N}  & 0 & \frac{1 - 2R_\nu}{4} &  -1 - 2R_\nu & -R_\nu & 0 & 0 \\
 0 & 0 & 0 & 0 & 0 & 0 & \frac{8}{5} & -2 & 0 & 0 \\
 0 & 0 & \frac{9}{4} \mathcal{M} \mathcal{N}\mathcal{W}  &  9 \mathcal{M}\mathcal{N}\mathcal{W}   & 0 & -\frac{9}{4}R_\nu \mathcal{M}\mathcal{W} & -9R_\nu\mathcal{M}\mathcal{W}  & 0 &  3\mathcal{W} & 9\mathcal{M}\mathcal{W} \\
 0 & 4\mathcal{B} & \mathcal{N}(1-\mathcal{B}) & 4\mathcal{N}(1-\mathcal{B})  & 0 & -R_\nu(1-\mathcal{B}) & -4R_\nu(1-\mathcal{B}) & -R_\nu & \frac{1}{w+1} & 1-8\mathcal{B}
\end{array}
\right),
\ee
where $\mathcal{N} = R_\nu -1$, $\mathcal{W}=w-1$, $\mathcal{M}=1+w-\beta$ and $\mathcal{B}=\frac{3\beta}{4(1+w)}$. Now, we can obtain the eigenvalues of $\mathbf{A}_0$ immediately,
\be
\lambda_i = \left\{-2,-1,0,0,0,0,- \frac{5}{2}   - \frac{\sqrt{1 - 32\,R_\nu
/5}}{2},- \frac{5}{2} + \frac{\sqrt{1 -  32\,R_\nu /5}}{2},
\lambda_{\rm{d}}^{-}, \lambda_{\rm{d}}^{+} \right\} \,,
\label{eq.eigenvals}
\ee
where
\be \lambda_{\rm{d}}^{\pm} =
\frac{-2 + 3\,w}{2}-\frac{3\beta}{1+w} \pm \frac{\sqrt{9w^4+30w^3+13w^2-12\beta w-28w+36\beta^2-12\beta-20}}{2(1+w)} \,. \label{eq:lambdas}
\ee

The approximate solutions to Eq.~(\ref{eq.dUdlnx_gen}) are a linear combination of $x^{\lambda_i}\bm{U}_0^{(i)}$, where $\bm{U}_0^{(i)}$ is the eigenvector corresponding to eigenvalue $\lambda_i$. Thus, the mode with negative $\rm{Re}(\lambda_i)$ will soon decay or oscillate while that with positive $\rm{Re}(\lambda_i)$ will dominate the evolution of each component. Under the condition that $0<R_\nu<0.405$, the only possible eigenvalues with positive $\rm{Re}(\lambda_i)$ are $\lambda_{\rm{d}}^{\pm}$. Thus, the sign of $\rm{Re}(\lambda_{\rm{d}}^{\pm})$ plays an important role in the evolutions of cosmological perturbations.

Let us first consider the $\rm{Re}(\lambda_{\rm{d}}^{\pm})<0$ case. Under this condition, the largest $\rm{Re}(\lambda_i)$ of Eq.~(\ref{eq.eigenvals}) is zero, which is fourfold degenerate. According to Ref.~\cite{Doran:2003xq}, the eigenvectors corresponding to these four eigenvalues construct the basis for the four initial conditions, one adiabatic mode and three isocurvature modes. Here we give the adiabatic initial conditions, obtained by setting the gauge invariant entropy perturbation $S_{AB}$ to be zero, where
\begin{equation} \label{eq.entropy}
S_{AB} = -3\h\frac{\rho_A}{\rho'_A} \Delta_A + 3\h \frac{\rho_B}{\rho'_B}\Delta_B.
\end{equation}
For DM, baryons, photons and neutrinos, the condition $S_{AB}=0$ gives
 \begin{equation} \label{eq.adiabatic_condition}
\Delta_{\rm{c}} = \Delta_{\rm{b}} = \frac{3}{4} \Delta_\gamma = \frac{3}{4}
\Delta_\nu \,.
 \end{equation}
Using the above equation we have
\begin{equation}
\label{eq.adiabatic_mode}
\bm{U}_0 = \left(
\begin{array}{c}
\Delta_{\rm{c}} \\ \tilde V_{\rm{c}}\\ \Delta_{\gamma} \\ \tilde V_{\gamma}
\\ \Delta_{\rm{b}} \\ \Delta_{\nu} \\ \tilde V_{\nu} \\ \tilde \Pi_{\nu}
\\ \Delta_{\rm{de}} \\
\tilde{V}_{\rm{de}}
\end{array}
\right) = C_1 \left(
\begin{array}{c}
3/4  \\
-(5/4) \mathcal{P} \\
1 \\
- (5/4) \mathcal{P}\\
3/4  \\
1 \\
- (5/4) \mathcal{P}\\
- \mathcal{P} \\
(3/4) \left( 1 + w-\beta \right) \\
- (5/4) \mathcal{P}
\end{array}
\right) \,,
\end{equation}
where $\mathcal{P} = \left( 15 + 4\,R_\nu \right)^{-1}$, and $C_1$
is a dimensionless normalization constant. We can see from Eq.~(\ref{eq.adiabatic_mode}) that DE automatically obeys the condition $S_{AB}=0$ and all the perturbations are stable. Thus, Eq.~(\ref{eq.adiabatic_mode}) give us the adiabatic initial conditions for stable $\rm{Re}(\lambda_{\rm{d}}^{\pm})<0$ case.

\begin{figure}[htbp]
  \includegraphics[width=8cm]{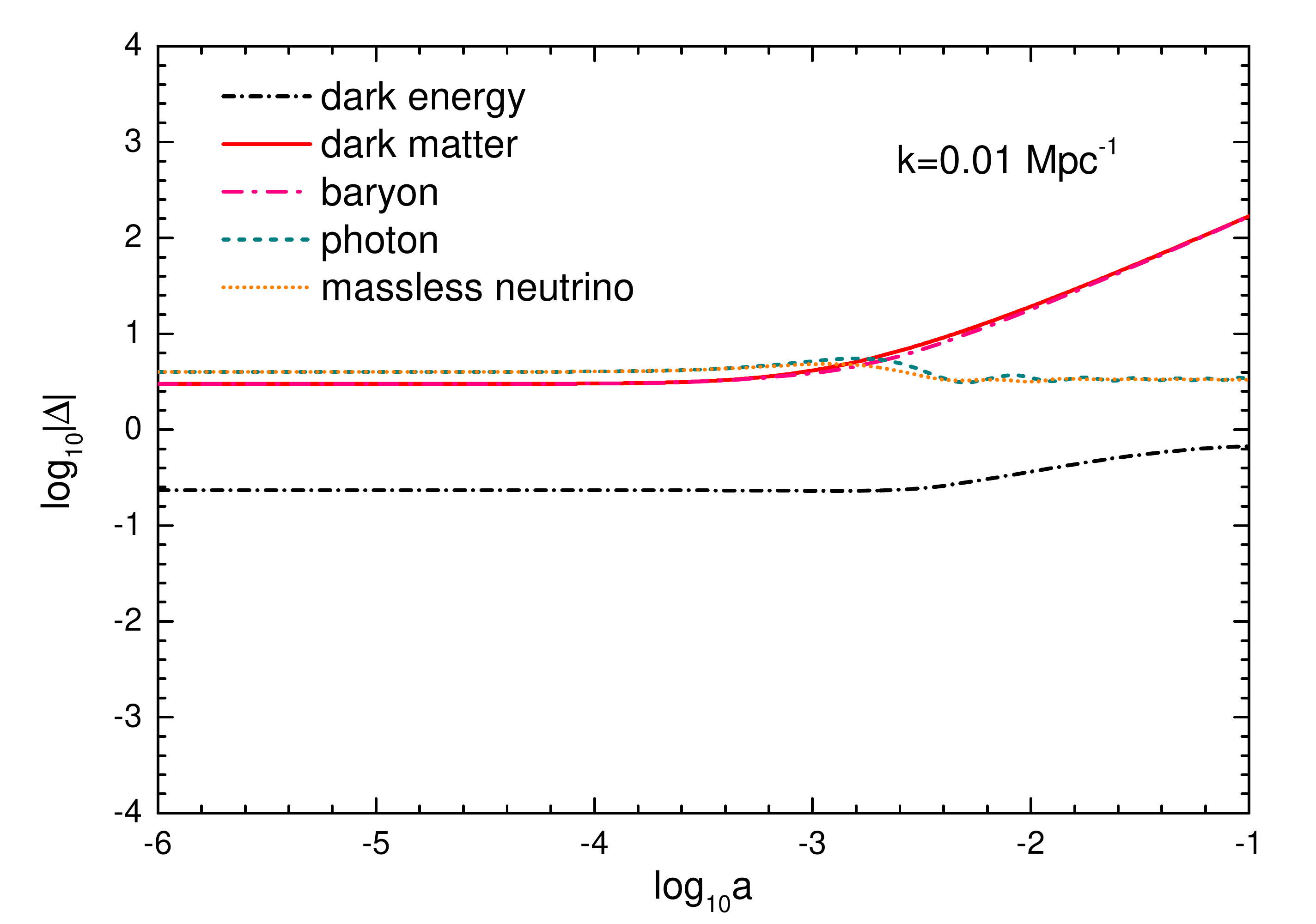}
  \includegraphics[width=8cm]{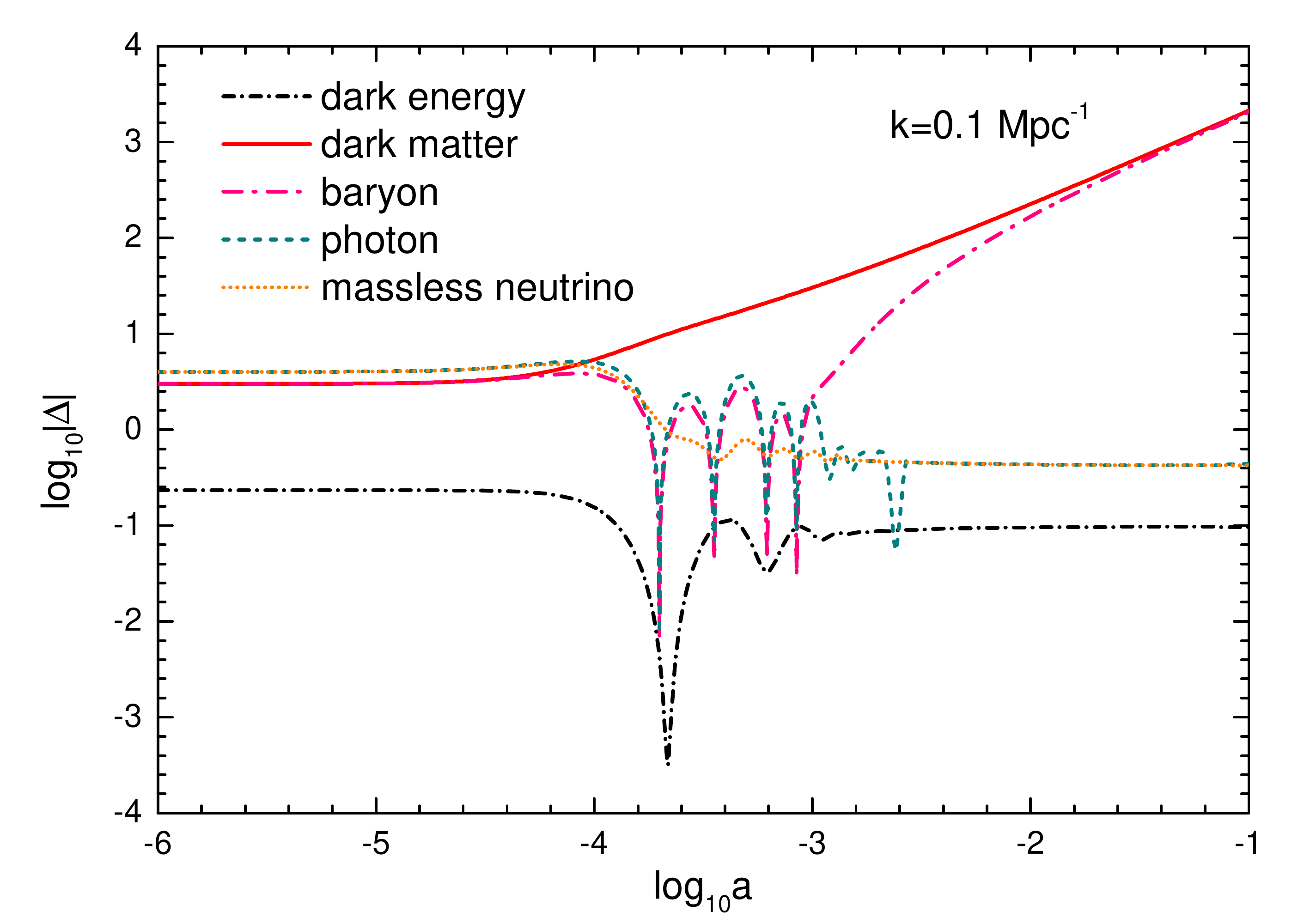}
  \includegraphics[width=8cm]{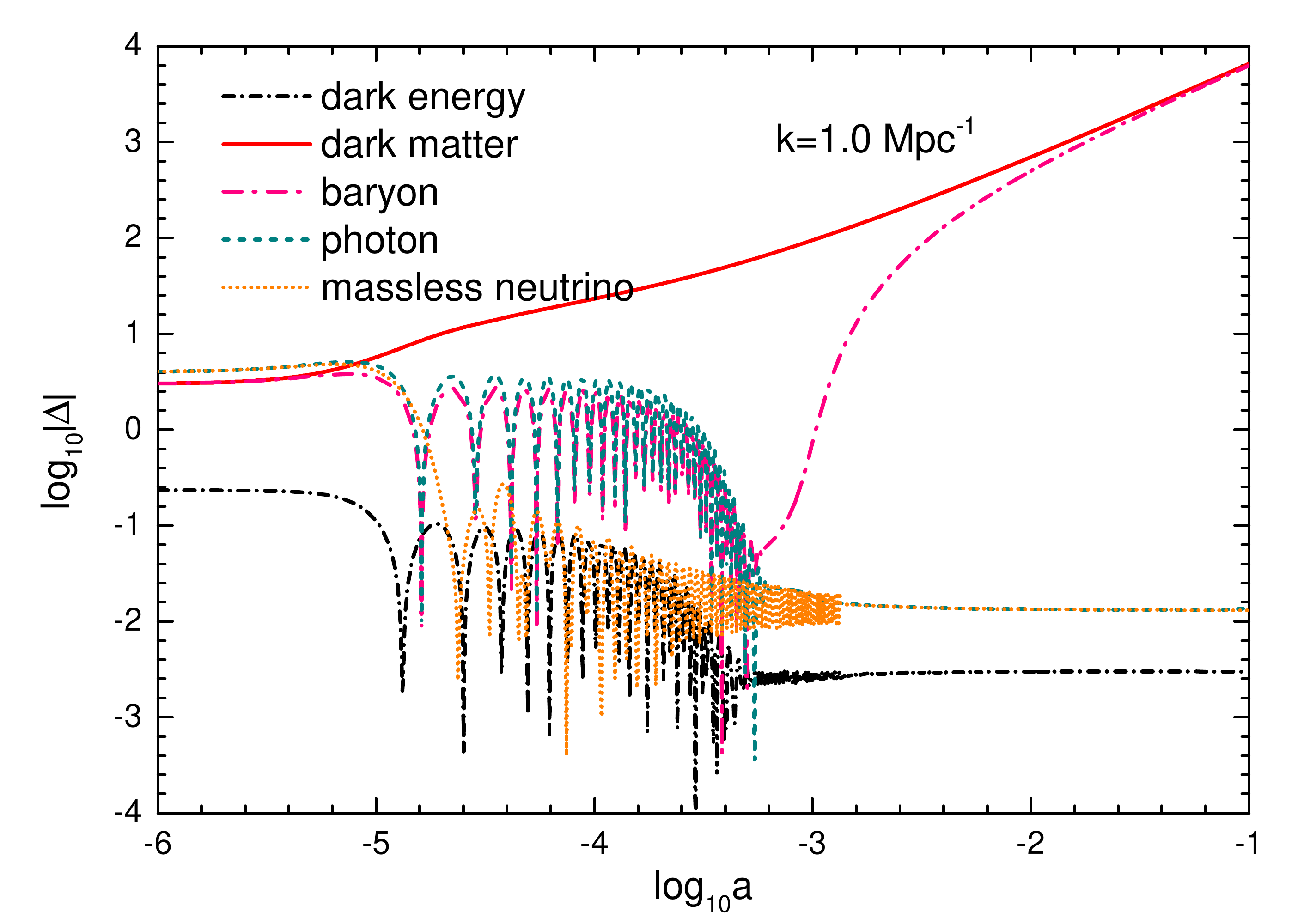}
  \includegraphics[width=8cm]{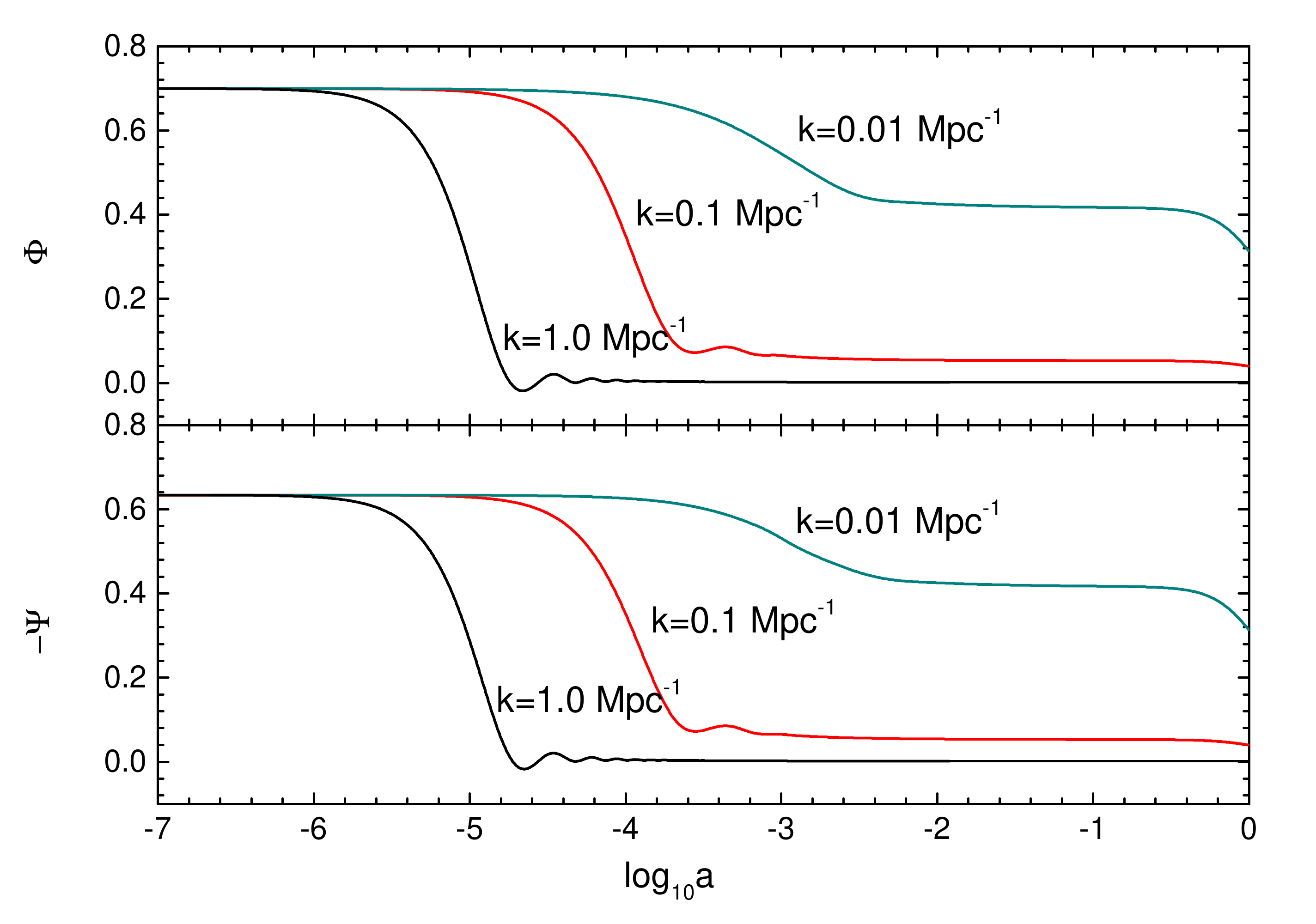}
  \caption{The evolutions of gauge invariant matter perturbations and metric perturbations for $k=0.01\,\rm{Mpc^{-1}}$, $k=0.1\,\rm{Mpc^{-1}}$ and $k=1.0\,\rm{Mpc^{-1}}$. Here, we choose $w=-0.98$ and $\beta=0.1$, and fix other cosmological parameters at the best-fit values from Planck.}\label{perturbationevolve}
\end{figure}

For the $\rm{Re}(\lambda_{\rm{d}}^{\pm})>0$ case, according to the discussion of Ref.~\cite{Majerotto:2009np}, the DE perturbation will dominate at the early times and drag other perturbations onto nonadiabatic blowup even if they are adiabatic at the initial times. Thus, $\rm{Re}(\lambda_{\rm{d}}^{\pm})>0$ corresponds to the instable case. From Eq.~(\ref{eq:lambdas}), we find that the parameter interval that can give stable cosmological perturbations ($\rm{Re}(\lambda_{\rm{d}}^{\pm})<0$) is $\beta>0$ under the assumption $w>-1$, which is the same as that in models $Q=3\beta H\rho_{\rm{de}}$ and $Q=\Gamma \rho_{\rm{de}}$. As an example for the stable case, we plot the evolutions of gauge invariant matter and metric perturbations in Fig.~\ref{perturbationevolve}, for $k=0.01\,\rm{Mpc^{-1}}$, $k=0.1\,\rm{Mpc^{-1}}$ and $k=1.0\,\rm{Mpc^{-1}}$. Here we choose $w=-0.98$, $\beta=0.1$, and fix other cosmological parameters at the best-fit values from Planck. We can clearly see that all the perturbation evolutions are normal and stable. Besides, we can also see from Fig.~\ref{perturbationevolve} that due to the existence of DE, the late-time evolutions of the metric perturbations $\Phi$ and $\Psi$ for $k=0.01\,\rm{Mpc^{-1}}$ suddenly change at $\log_{10}a\simeq-0.4$, which is the source of the late-time integrated Sachs-Wolfe effect \cite{ISW} on the large scale. Figure~\ref{perturbationevolve} also presents an exotic feature that the perturbation of DE oscillates when baryons and photons are tightly coupled. This oscillation feature for DE arises from the $\mathcal{K}$ term in Eq.~(\ref{eq.delta'de_ourQ}), since $\mathcal{K}$ contains the total velocity $\theta$ which oscillates when baryons and photons are tightly coupled. However, as pointed out in Ref.~\cite{Gavela:2010tm}, the $\mathcal{K}$ term does not significantly affect the observational constraint results.

\section{Observational constraints}\label{sec:constraint}

In this section, we constrain our interacting dark energy model using current observational data. What we most care about here is whether a positive $\beta$ required by stable cosmological perturbations is consistent with the observations. We modify the {\tt CAMB} code \cite{camb} for our interacting dark energy model. In the synchronous gauge ($\phi=B=0$, $\psi=\eta$, and $k^2E=-\frac{h}{2}-3\eta$), Eqs.~(\ref{eq.delta'de_ourQ})--(\ref{eq.theta'c_ourQ}) become
 \begin{eqnarray}
&&\delta_{\rm{de}}'=-3{\cal H}(1-w)\big[\delta_{\rm{de}}+3{\cal
H}(1+w){\theta_{\rm{de}} \over k^2}\big]
-(1+w)\theta_{\rm{de}}- (1+w)\frac{h'}{2}
\nonumber\\
&&~~~~+3\beta\mathcal{H}R_{\rm{c}}\left[(1-R_{\rm{c}})(\delta_{\rm{c}}-\delta_{\rm{de}})
+\frac{\theta}{3\mathcal{H}}+\frac{h'}{6\mathcal{H}}+ 3\mathcal{H} (1-w)\frac{\theta_{\rm{de}}}{k^2}\right],\\
&& \theta_{\rm{de}}'=2{\cal H}\theta_{\rm{de}}+ {k^2\delta_{\rm{de}}
\over (1+w)}-{3\beta\mathcal{H}\over{1+w}}R_{\rm{c}}(2\theta_{\rm{de}}-\theta_{\rm{c}}),\\
&&\delta_{\rm{c}}'=-\theta_{\rm{c}}-\frac{h'}{2} -3\beta\mathcal{H}(1-R_{\rm{c}})\left[R_{\rm{c}}(\delta_{\rm{de}}-\delta_{\rm{c}})+\frac{\theta}{3\mathcal{H}}+\frac{h'}{6\mathcal{H}}\right],\\
&& \theta_{\rm{c}}'=-{\cal H}\theta_{\rm{c}}.
 \end{eqnarray}
We use the adiabatic initial conditions obtained in the last section to solve the cosmological perturbation equations, and set $\theta_{\rm{c}}=0$ at the initial times so that DM is always at rest in the synchronous gauge.

We use the public Markov-Chain Monte-Carlo (MCMC) package {\tt CosmoMC} \cite{cosmomc} to explore the space of the cosmological parameters. The free parameter vector is
\begin{equation}
\label{parameter} \left\{\Omega_{\rm{b0}}h^2,\,\Omega_{\rm{c0}}h^2,\,H_0,\,
\tau,\, w,\, \beta, n_{\rm{s}},\, {\rm{ln}}(10^{10}A_{\rm{s}})\right\}.
\end{equation}
Here, $h$ is the Hubble constant $H_0$ in units of 100 ${\rm km\,s^{-1}\,Mpc^{-1}}$, $\tau$ is the optical depth to
reionization, and ${\rm{ln}}(10^{10}A_{\rm{s}})$ and $n_{\rm{s}}$ are the amplitude and
the spectral index of the primordial scalar perturbation power
spectrum for the pivot scale $k_0=0.05\,\rm{Mpc}^{-1}$. The priors of all the free parameters used in running MCMC are listed in Table~\ref{tab.params}. Note that we directly use $H_0$ as a free parameter in place of the commonly used parameter $\theta_{\rm{MC}}$ defined as the approximation to the ratio of the comoving sound horizon at $z=z_{\ast}$ (with $z_{\ast}$ the redshift when the optical depth equals unity). {\tt CosmoMC} using $\theta_{\rm{MC}}$ instead of $H_0$ is due to that $\theta_{\rm{MC}}$ is much better constrained than $H_0$. However, the value of $z_{\ast}$ used to derive $\theta_{\rm{MC}}$ comes from a fitting formula in Ref.~\cite{Hu:1995en}, which assumes a standard noninteracting background evolution. In our work, we fix the effective number of neutrinos $N_{\rm{eff}}=3.046$ and the total mass of standard neutrinos $\Sigma m_{\nu}=0.06\,\rm{eV}$, adopted as the same as Ref.~\cite{Ade:2013zuv}.

For the observations, we use the following data sets:
 \begin{itemize}
 \item
 The cosmic microwave background (CMB) observations including the high-$l$ TT likelihood at $l=50$--2500 and the low-$l$ TT likelihood at $l<50$ from Planck and low-$l$ TE, EE, BB likelihood (polarization measurements) from 9-year WMAP. All the data can be downloaded from Planck Collaboration \cite{planckdata}.
\item
The type Ia supernova (SN) observations of 580 data from Union2.1 sample (without considering the systematic errors)~\cite{Suzuki:2011hu}.
\item
The baryon acoustic oscillation (BAO) data at $z=0.106$ from the 6dF Galaxy Survey \cite{Beutler:2011hx}, $z=0.35$ from the SDSS DR7 measurement \cite{Padmanabhan:2012hf} and $z=0.57$ from BOSS DR9 measurement \cite{Anderson:2012sa}.
\item
The Hubble constant measurement, $H_0=73.8\pm 2.4\,{\rm km\,s^{-1}\,Mpc^{-1}}$,
from the HST \cite{riess2011}.
\end{itemize}

\begin{figure}[htbp]
  \includegraphics[width=17cm]{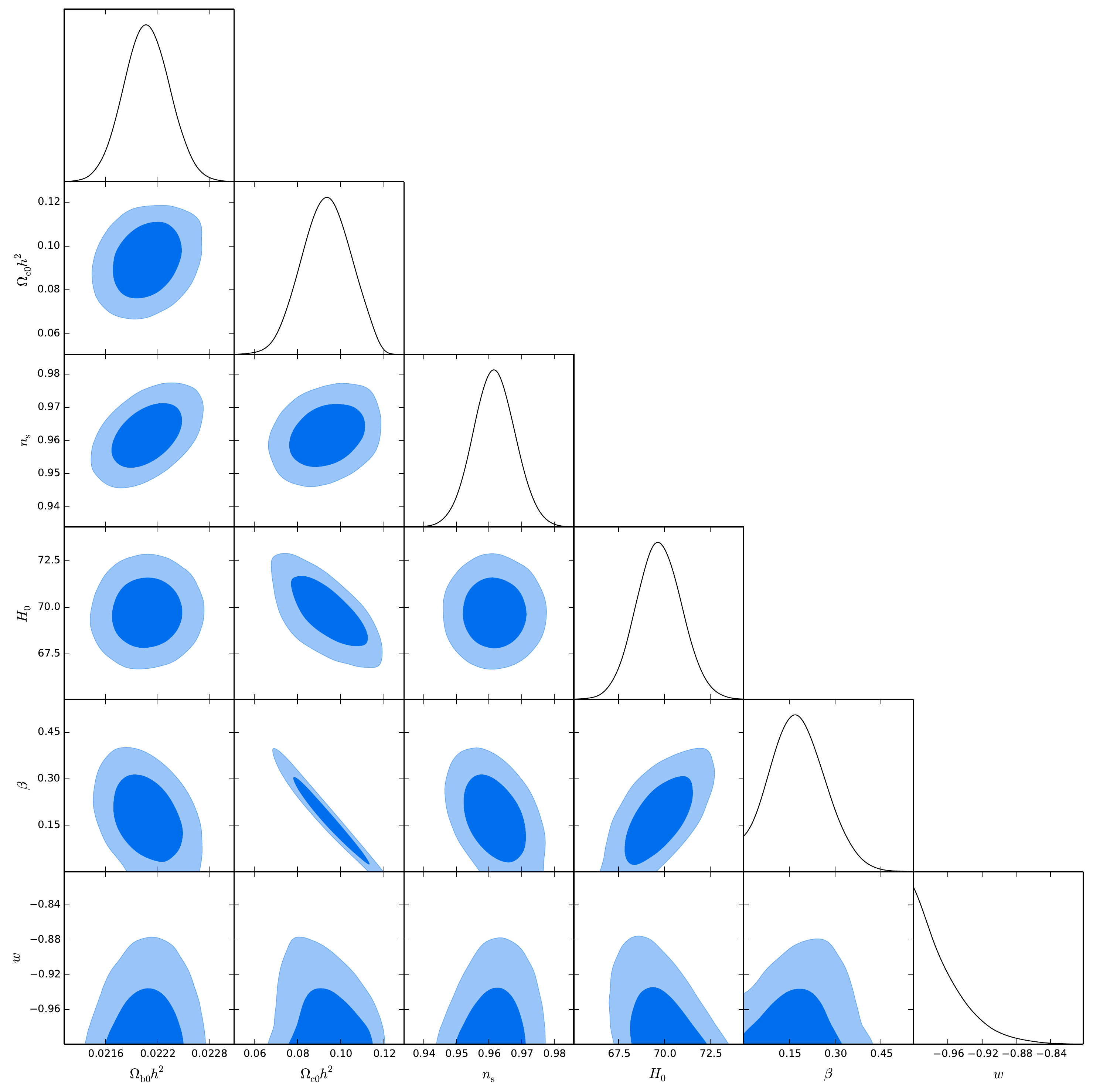}
  \caption{The one-dimensional marginalized distributions and two-dimensional marginalized 68\% and 95\%
contours, for the parameters in our interacting dark energy model.}\label{fig.fitresult}
\end{figure}

Our fit results are summarized in Table \ref{tab.params} and Fig.~\ref{fig.fitresult}. The best fit of the coupling constant $\beta$ is 0.1385 and its $68\%$ limits are $0.178^{+0.081}_{-0.097}$, which are greater than 0 at more than 1$\sigma$ confidence level. This result is consistent with that obtained in a latest fit work \cite{Salvatelli:2013wra} using Planck data to constrain the $Q=3\beta H\rho_{\rm{de}}$ model. From Fig.~\ref{fig.fitresult}, we find that there exists a strong anticorrelation between the coupling constant $\beta$ and the physical cold dark matter density $\Omega_{\rm{c0}}h^2$, which results in a low value of $\Omega_{\rm{m0}}$ and a high value of $\Omega_{\rm{de0}}$ as shown in Table \ref{tab.params}, since a positive coupling constant $\beta$ is favored by observations. These results can be easily understood. For our interacting dark energy model, a positive coupling constant $\beta$ leads to the energy transfer from DM to DE, and so the stronger coupling is, the lower energy density of matter becomes. 
\begin{table}[htbp]\caption{The fit results for the free parameters and some derived parameters. We give their best-fit values as well as the marginalized 68\% confidence limits. We also present the prior ranges of the free parameters used in running MCMC.}\label{tab.params}
\begin{tabular}{lccc}
\hline\hline
Parameter & Prior & Best fit & 68\% limits \\
\hline
$\Omega_{\rm{b0}}h^2$&[0.005, 0.1] &0.02208&$0.02208^{+0.00025}_{-0.00025}$\\
$\Omega_{\rm{c0}}h^2$& [0.001, 0.99]&0.0987&$0.0934^{+0.0110}_{-0.0109}$\\
$H_0$& [20, 100]&70.0&$69.7^{+1.2}_{-1.2}$\\
$\tau$& [0.01, 0.8]&0.082&$0.089^{+0.012}_{-0.014}$\\
$w$& [$-1$, $-0.3$]&$-0.9908$&$-0.9657^{+0.0071}_{-0.0342}$\\
$\beta$ & [0, 1.0]&0.1385&$0.178^{+0.081}_{-0.097}$\\
$n_{\rm{s}}$& [0.9, 1.1]&0.9630&$0.9616^{+0.0062}_{-0.0062}$\\
${\rm{ln}}(10^{10}A_{\rm{s}})$& [2.7, 4.0]&3.074&$3.087^{+0.025}_{-0.025}$\\
\hline
$\Omega_{\rm{de0}}$&&0.7521&$0.7603^{+0.0286}_{-0.0290}$\\
$\Omega_{\rm{m0}}$&&0.2479&$0.2397^{+0.0290}_{-0.0286}$\\
$z_{\rm{re}}$&&10.42&$10.00^{+1.09}_{-1.09}$\\
${\rm{Age}}/{\rm{Gyr}}$&&13.744&$13.755^{+0.038}_{-0.038}$\\
$100\theta_*$&&1.04184&$1.04151^{+0.00058}_{-0.00058}$\\
\hline
\end{tabular}
\end{table}

\section{Conclusions}

There exists an important possibility that dark energy interacts with cold dark matter in some direct, nongravitational way. In this paper, we focus on the interacting dark energy models with constant $w$ (and $w>-1$); this class of models may also be called interacting $w$CDM model. For the widely studied forms of interaction, $Q\propto\rho_{\rm c}$ (or $Q\propto H\rho_{\rm c}$) and $Q\propto\rho_{\rm de}$ (or $Q\propto H\rho_{\rm de}$), there are some fatal flaws in the model. For instance, $Q\propto\rho_{\rm c}$ leads to a large-scale instability at the early times, and $Q\propto\rho_{\rm de}$ (with a positive coupling) gives rise to an unphysical result for the evolution of cold dark matter density, i.e., negative $\rho_{\rm c}$ in the future. In order to overcome these flaws, we propose in this paper an interacting $w$CDM model with $Q=3\beta H \frac{\rho_{\rm{de}}\rho_{\rm{c}}}{\rho_{\rm{de}}+\rho_{\rm{c}}}$, and show that this model is a reasonable, large-scale stable interacting dark energy model.

By carefully designing the form of $Q$, this model gets excellent features: At early times, $Q\propto\rho_{\rm de}$, and so the early-time large-scale instability can be avoided; in the future, $Q\propto\rho_{\rm c}$, and thus the problem of negative $\rho_{\rm c}$ can be eliminated.

We have calculated the cosmological perturbations in this model. We also considered the perturbation of the Hubble parameter $H$ in the calculation in order to get the gauge invariant equations for the dark matter and dark energy perturbations. We find that the cosmological perturbations in this interacting $w$CDM model (with $w>-1$) are stable during the whole expansion history provided that $\beta>0$. We have also used the CMB temperature data from Planck and CMB polarization data from 9-yr WMAP, in conjunction with the SN data, BAO data, and $H_0$ measurement, to place stringent constraints on this model (with eight parameters). We find support for $\beta>0$ from the joint constraint: $0.081<\beta<0.259$ (68\% CL).

We also show that this interacting $w$CDM model is, actually, equivalent to the decomposed NGCG model~\cite{Zhang:2004gc}, with the relation $\beta=-\alpha w$. The excellent theoretical features and the support from observations all indicate that the decomposed NGCG model should be payed more attentions in the future. Recently, the Planck Collaboration reported that the CMB data are in tension with other astrophysical data sets such as the direct measurement of $H_0$ and the SN data, based on the 6-parameter $\Lambda$CDM model. It has been found that the tension between CMB and $H_0$ could be greatly reduced if a dynamical dark energy is considered (e.g., $w$CDM model or holographic dark energy model)~~\cite{Li:2013dha}. Furthermore, if the possible evolution of the color-luminosity parameter in SN is also considered, the tension between CMB and SN might also be significantly reduced~\cite{Wang:2013a,Wang:2013b}. Therefore, it is of great interest to see if the Planck data and other astrophysical data are consistent with each other in the framework of interacting dark energy. We will leave the full analysis on this model by using the observational data in the future work.

\begin{acknowledgments}
We acknowledge the use of {\tt CosmoMC}. This work is supported by the National Natural Science Foundation of China (Grants No. 10975032 and No. 11175042) and by the National Ministry of Education of China (Grants No. NCET-09-0276 and No. N120505003).
\end{acknowledgments}

\appendix
\section{Decomposed NGCG model}\label{appendixNGCG}

The EOS of the NGCG is given by \cite{Zhang:2004gc}
\begin{equation}
p_{\rm Ch} = - {\tilde{A}(a) \over \rho_{\rm Ch}^\alpha},
\label{eqstate}
\end{equation}
where $\tilde{A}(a)=-wAa^{-3(1+w)(1+\alpha)}$,
with $\alpha$ a dimensionless parameter and $A$ a positive constant.
The NGCG is designed as a unification scheme for DE and DM; however,
on the other hand, it can also be viewed as an interacting $w$CDM model, provided that
it is decomposed into the two components, DE (with constant $w$) and CDM,
\begin{equation}\label{eq:decompgcg}
\rho_{\rm Ch}=\rho_{\rm{de}}+\rho_{\rm{c}}.
\end{equation}
The continuity equations for DE and DM are given by Eqs.~(\ref{rhodedot}) and (\ref{rhocdot}).
Since DM is pressureless, the pressure of the NGCG is provided only by DE, i.e., $p_{\rm{Ch}}=p_{\rm{de}}$.
Therefore, from Eqs.~(\ref{eqstate}) and (\ref{eq:decompgcg}), we have
\begin{equation}\label{AA}
A=\rho_{\rm{de}}(\rho_{\rm{de}}+\rho_{\rm{c}})^\alpha a^{3(1+w)(1+\alpha)}.
\end{equation}
Since $A$ is a constant, we have $\dot{A}=0$.
Furthermore, using $\dot{A}=0$ and Eqs.~(\ref{rhodedot}), (\ref{rhocdot}) and (\ref{AA}), we obtain the
interaction term,
\begin{equation}\label{eq:QNGCG}
 Q=-3\alpha wH\rho_{\rm{de}}R_{\rm{c}}.
\end{equation}
Comparing Eq.~(\ref{eq:QNGCG}) with Eq.~(\ref{interaction}), we find the relation $\beta=-\alpha w$.
So, the interacting $w$CDM model with such an interaction term is actually equivalent to the decomposed
NGCG model.


\setcounter{figure}{0}
\renewcommand{\thefigure}{B\arabic{figure}}
\section{Model with the energy-momentum transfer parallel to the four-velocity of dark energy}\label{appendixDEV}
\begin{figure}[!htbp]
  \includegraphics[width=8cm]{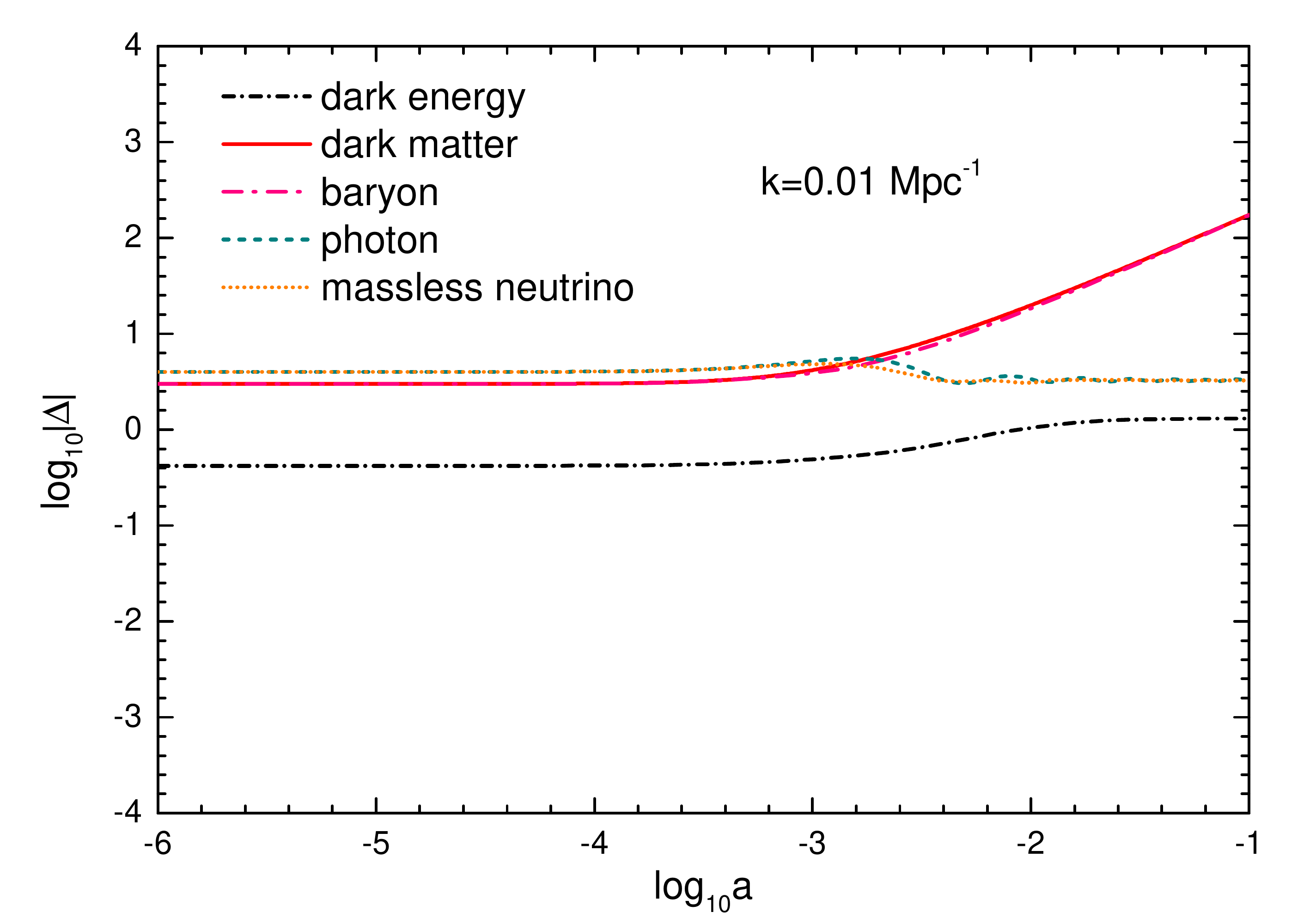}
  \includegraphics[width=8cm]{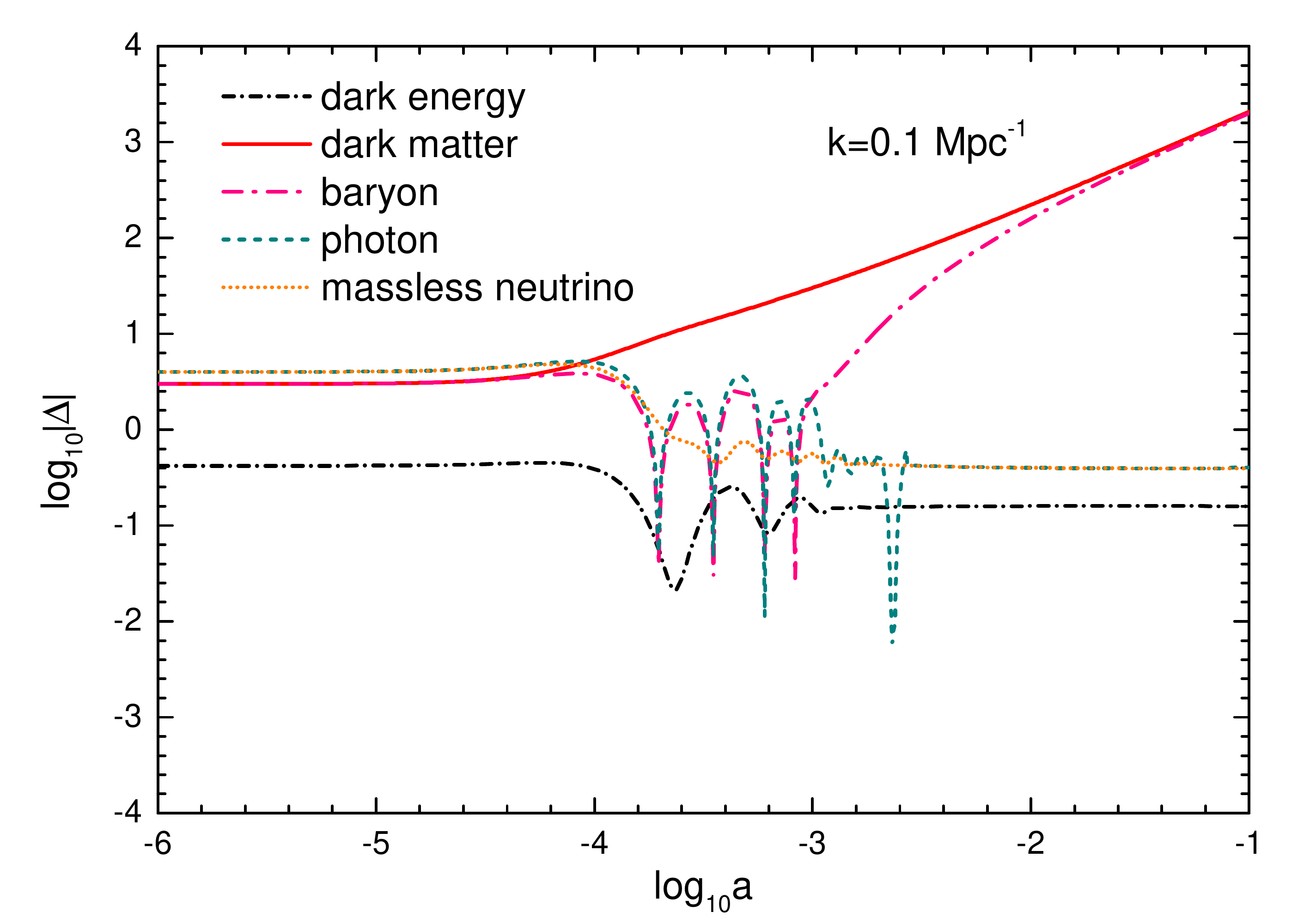}
  \includegraphics[width=8cm]{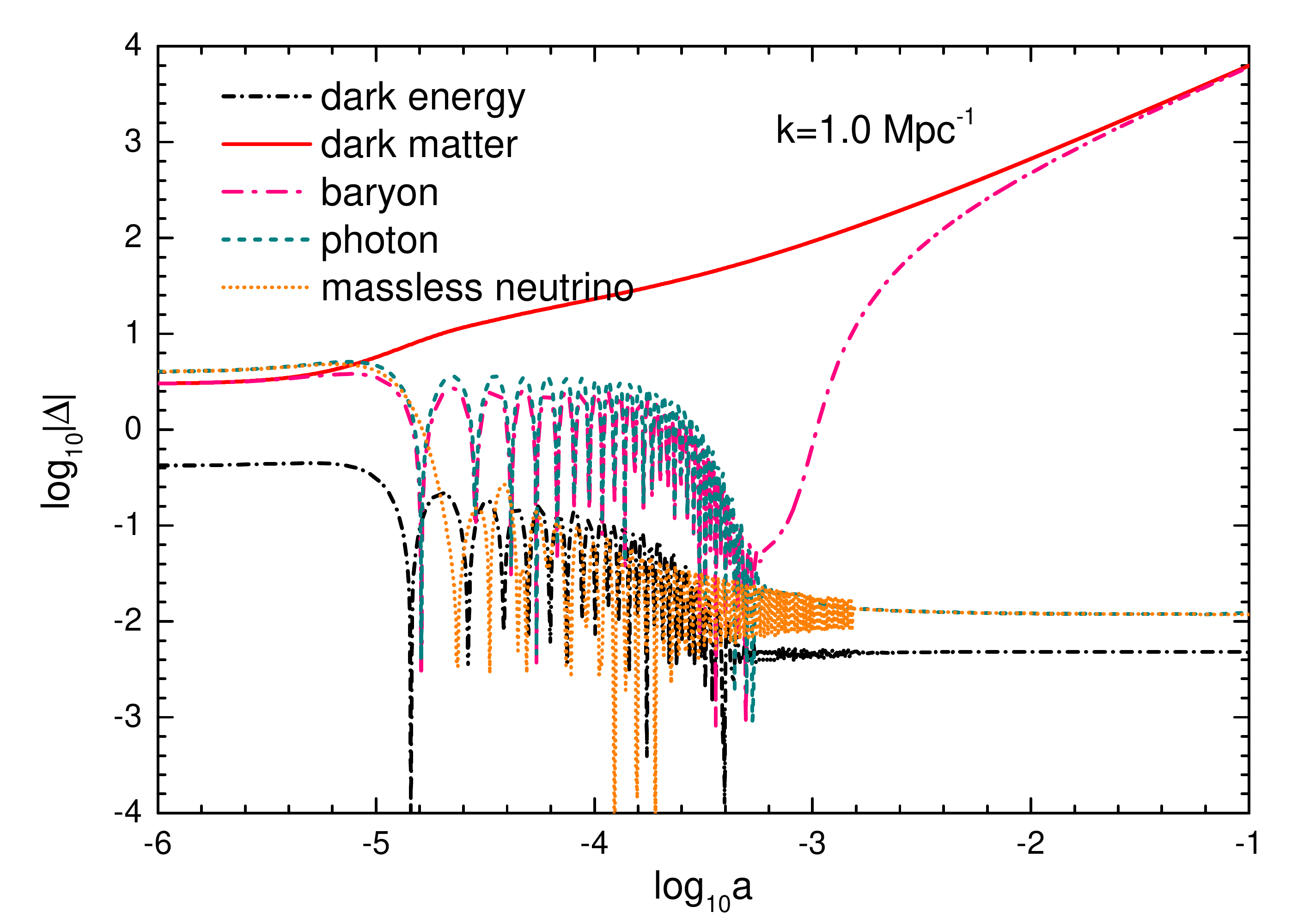}
  \includegraphics[width=8cm]{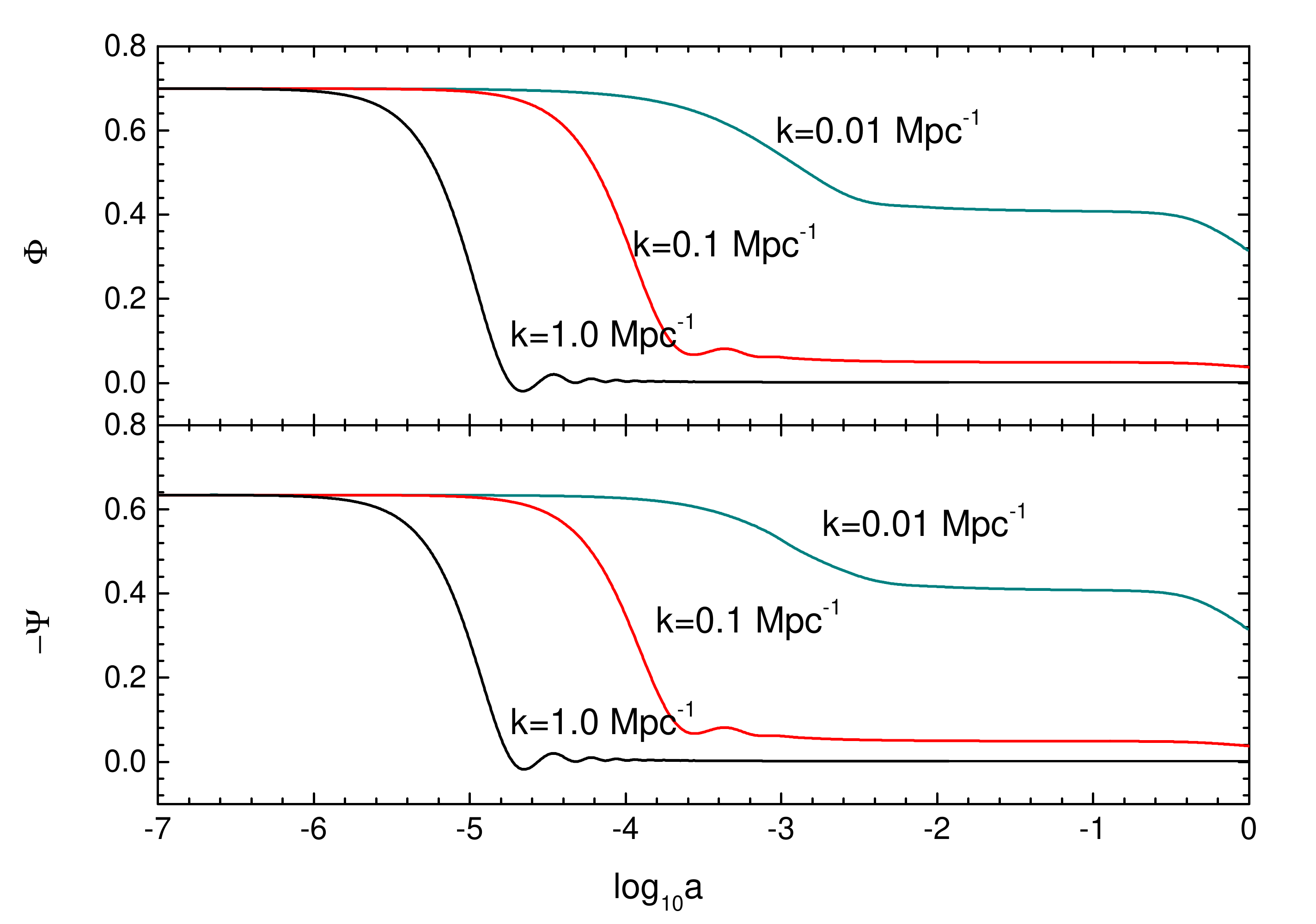}
  \caption{The evolutions of gauge invariant matter perturbations and metric perturbations for $k=0.01\,\rm{Mpc^{-1}}$, $k=0.1\,\rm{Mpc^{-1}}$ and $k=1.0\,\rm{Mpc^{-1}}$ in the case with the energy-momentum transfer parallel to the four-velocity of DE. Here, all the values of the cosmological parameters are set as the same as those used in Fig.~\ref{perturbationevolve}.}\label{perturbationevolve_de}
\end{figure}

The stability of the cosmological perturbations is independent of the choice of the energy-momentum transfer type and the cosmological constraint results are also similar for different types of the energy-momentum transfer \cite{Clemson:2011an}. Here we give a brief discussion on the evolutions of the cosmological perturbations when the energy-momentum transfer is parallel to the four-velocity of DE. In this case, the covariant interaction form is
\begin{equation}
aQ_{\rm{c}}^\mu=-aQ^\mu_{\rm{de}}=-3\beta\mathcal{H} \rho_{\rm{de}}R_{\rm{c}} u_{\rm{de}}^\mu.
\end{equation}
This covariant interaction form gives the same energy balance equations (\ref{eq.delta'de_ourQ}) and (\ref{eq.delta'c_ourQ}) for DE and DM but different momentum balance equations,
\bea
&& \theta'_{\rm{de}} -2 \mathcal H\theta_{\rm{de}}
-\frac{k^2}{(1+w)}\delta_{\rm{de}} - k^2\phi =
-\frac{3\beta\mathcal{H}}{1+w}R_{\rm{c}}
\theta_{\rm{de}}\,,\label{eq.theta'de_de} \\
&& \theta'_{\rm{c}} + \mathcal H \theta_{\rm{c}} -k^2\phi = 3\beta\mathcal{H}(1-R_{\rm{c}})
(\theta_{\rm{c}}-\theta_{\rm{de}})\,.\label{eq.theta'c_de}
 \eea
Using Eqs.~(\ref{eq.theta'de_de}) and (\ref{eq.theta'c_de}), we can derive the evolutions of gauge invariant matter and metric perturbations, plotted in Fig.~\ref{perturbationevolve_de}, for $k=0.01\,\rm{Mpc^{-1}}$, $k=0.1\,\rm{Mpc^{-1}}$ and $k=1.0\,\rm{Mpc^{-1}}$. Here all the values of the cosmological parameters are set as the same as those used in Fig.~\ref{perturbationevolve}. We can clearly see that the perturbations are also stable in the case with the energy-momentum transfer parallel to the four-velocity of DE.

\end{document}